\documentclass[conference]{IEEEtran}
\IEEEoverridecommandlockouts
\usepackage{cite}
\usepackage{amsmath,amssymb,amsfonts}
\usepackage[hyphens]{url}
\usepackage[hidelinks]{hyperref}
\hypersetup{breaklinks=true}
\usepackage{algorithmic}
\usepackage{graphicx}
\usepackage{textcomp}
\usepackage{color,soul}
\usepackage{booktabs}
\usepackage[table,xcdraw]{xcolor}

\usepackage{caption, floatrow}

\def\BibTeX{{\rm B\kern-.05em{\sc i\kern-.025em b}\kern-.08em
    T\kern-.1667em\lower.7ex\hbox{E}\kern-.125emX}}
\begin{document}


\title{Mass Adoption of NATs: Survey and experiments on carrier-grade NATs}

\author{\IEEEauthorblockN{Orestis Kanaris}
\IEEEauthorblockA{\textit{Delft University of Technology}\\
Delft, Netherlands \\
O.Kanaris@student.tudelft.nl}



\and

\IEEEauthorblockN{Johan Pouwelse (msc supervisor)}
\IEEEauthorblockA{\textit{Delft University of Technology}\\
Delft, Netherlands \\
J.A.Pouwelse@tudelft.nl}


}

\maketitle

\begin{abstract}
In recent times, the prevalence of home NATs and the widespread implementation of Carrier-Grade NATs have posed significant challenges to various applications, particularly those relying on Peer-to-Peer communication. This paper addresses these issues by conducting a thorough review of related literature and exploring potential techniques to mitigate the problems. The literature review focuses on the disruptive effects of home NATs and CGNATs on application performance. Additionally, the study examines existing approaches used to alleviate these disruptions. Furthermore, this paper presents a comprehensive guide on how to puncture a NAT and facilitate direct communication between two peers behind any type of NAT. The techniques outlined in the guide are rigorously tested using a simple application running the IPv8 network overlay, along with their built-in NAT penetration procedures. To evaluate the effectiveness of the proposed techniques, 5G communication is established between two phones using four different Dutch telephone carriers. The results indicate successful cross-connectivity with three out of the four carriers tested, showcasing the practical applicability of the suggested methods.
\end{abstract}


\section{Introduction}
Internet connectivity has become a fundamental necessity. With the recent skyrocket in internet-connected devices, the demand for internet access keeps increasing, leading to the inevitable shortage of IPv4 addresses. This was anticipated since the late 1980s leading to the design of IPv6 \cite{murphy2005ipv6}. However, with the limited supply of IPv4 addresses and the rate of the rollout of IPv6 being slower than the depletion of IPv4, ISPs have had to find ways to conserve their address space.\par

Carrier-grade NATs (CGNATs) have emerged as a solution to address this issue. CGNATs allow ISPs to share a single IP address with individual devices on a local network and translate them to a single public IP address for communication with the broader internet.\par

While CGNAT has been successful in maximizing IPv4 address usage, it has several implications, particularly concerning peer-to-peer (P2P) protocols. P2P protocols rely on direct connections between peers, but with CGNAT, direct connections are not possible as all traffic must be routed through the carrier's NAT device. This can lead to increased latency, slower download speeds and reduced overall performance. According to Yangyang Liu et al., \cite{LIU2014197} when exchanging files on BitTorrent, peers behind NATs tend not to get favoured thus significantly decreasing their download speed and in some cases even degenerating the download into a client-server interaction \cite{LIU2014197}. There have been multiple suggested ways of bypassing a NAT to establish a direct P2P connection but the most prominent and easiest to implement is NAT puncturing where one can establish a direct channel of communication with another peer given that they can find out the receiver's IP address \cite{halkes_pouwelse_2011, ford2005peer, ietf_draft_trustchain, guha2005characterization}. This is also apparent in P2P networks involving mobile phones in cellular networks which are by default behind a NAT \cite{9794301}.\par

Dutch telecommunication providers have also implemented CGNAT to conserve their IPv4 address space. The Netherlands has a relatively high internet usage \cite{Kemp_2022}, leading to a fast shortage of IPv4 addresses. The implementation of CGNAT allows Dutch ISPs to share a single IPv4 address among multiple users, thereby conserving the available address space \cite{skala2020technology}.\par

However, the implementation of CGNAT has not been without controversies. Privacy advocates argue that CGNAT undermines users' privacy by making it difficult to track individual users' online activity since with CGNAT, all users appear to have the same IP address \cite{europol_2017}. Additionally, CGNAT makes it more challenging for users to establish secure and direct connections with other users, potentially exposing their private data to more public networks.\par 


This work examines the need and the technology behind NATs, specifically CGNATs and how they affect P2P applications. Then a simple app which reproduces the NAT penetration algorithms of the main literature developed for the scope of this paper will be evaluated by attempting to communicate between various Android phones which use 5G from different carriers operating in the Netherlands.\par

The subsequent sections explore the impact of NATs on P2P protocols, techniques for penetrating NATs, and the reproducibility of results obtained from the literature. Additionally, this study examines the legal implications associated with Carrier-Grade NATs and proposes an alternative solution that surpasses the limitations of traditional Carrier-Grade NATs. Finally, a comprehensive discussion brings together the findings and insights gained throughout this research. By delving into these sections, we aim to enhance our understanding of P2P network connections, address the challenges imposed by NATs, and offer potential solutions for a more efficient and scalable network architecture. A brief overview of all the alternatives and workarounds to the problems introduced by NATs reviewed in this paper can be found in table \ref{tab:taxonomy}.\par

\begin{table*}[t]
    \begin{tabular}{|r|p{1.5in}|p{4.5in}|p{.57in}|}
    \hline

    \textbf{Year} & \textbf{Technique} & \textbf{Description} & \textbf{Literature}\\\hline
    
    2003 & NAT Puncturing & A technique that allows direct communication between two devices behind separate NAT routers by utilizing temporary port mappings or establishing a relay server to facilitate communication.  & \cite{halkes_pouwelse_2011, ford2005peer, ietf_draft_trustchain, guha2005characterization, rfc3489} \\\hline

    2008 & Extending IPv4 Address Space & Extending the IPv4  address space by "stealing" 6 bits from the port number in TCP/UDP thus extending the address space by 10 bits & \cite{maennel2008better}\\\hline

    2010 & pwnat & Autonomous NAT traversal using fake ICMP messages to initially contact the NATed peer without using any third party server& \cite{muller2010autonomous}  \\\hline
    
    2011 & Dual Stack-Lite & An IPv6 transition mechanism that enables the coexistence of IPv4 and IPv6 by encapsulating IPv4 packets within IPv6, allowing service providers to conserve IPv4 addresses while migrating to IPv6 & \cite{rfc7021, rfc6333}  \\\hline

    2013 & Universal Plug`n`Play Internet Gateway Device (UPnP IGD) & Enables devices on a network to automatically set up and manage port forwarding, allowing for seamless communication and connectivity. & \cite{rfc6970} \\\hline
    
    2013 & NAT Port Mapping Protocol (NAT-PMP) & Simplifies the process of port mapping by allowing devices to automatically request and configure port mappings on a network's NAT router. & \cite{rfc6886} \\\hline
    
    2013 & Port Control Protocol (PCP)  & Similar to NAT-PMP but supports IPv6 too & \cite{rfc6887} \\\hline
    
    2018 & Interactive Connectivity Establishment (ICE) & Facilitates the establishment of peer-to-peer connections across networks by dynamically selecting and negotiating the optimal communication path & \cite{rfc8445} \\\hline
    
    2019 & Birthday-Paradox based NAT Puncturing  & NAT Puncturing but exploits the birthday paradox to figure out the address port mapping on hard NATs & \cite{swadling2019birthday, anderson_2020, anderson_2022, danderson} \\\hline
    
    2021 & QUIC Penetration   & If TCP is necessary due to a stream-oriented connection, switch to QUIC and then adapt the UDP-puncturing techniques to work with QUIC & \cite{quic} \\\hline
    
    \end{tabular}\\
    \caption{Overview of all peer-to-peer techniques to establish communication behind NATs order by year of initial proposal}
    \label{tab:taxonomy}
\end{table*}

\section{Peer-to-Peer Network Connections}

In the context of this essay, it is crucial to first understand what a UDP session is and what an incoming connection is ---in the Peer-to-Peer context. This tutorial provides a concise overview of these concepts, explaining what they entail and how they function within UDP-based peer-to-peer networks.

UDP is a lightweight, connectionless transport protocol within the Internet Protocol (IP) suite. In UDP peer-to-peer connectivity, a UDP session refers to the exchange of data between two peers without the need for a persistent connection. Key characteristics of UDP sessions in peer-to-peer networks are as follows:

\texttt{Connectionless Communication:}
UDP sessions operate without establishing a dedicated connection between peers. Instead, UDP sends independent datagrams, each containing the necessary information to reach its destination. Peers can initiate data transmission without prior handshaking or negotiation.

\texttt{Unreliable Delivery:}
Unlike protocols like TCP (Transmission Control Protocol), UDP does not provide built-in error correction or retransmission of lost packets. Consequently, UDP sessions offer unreliable delivery, meaning the protocol does not guarantee that every packet will be received. It is the responsibility of the receiving application to handle any packet loss or errors.

\texttt{Datagram Structure:}
UDP transmits data in discrete units called datagrams. Each datagram contains a source and destination port number, along with payload data. The size of the payload is limited by the maximum transmission unit (MTU) of the network. Peers can exchange these datagrams freely, allowing for quick and lightweight communication.\par

\textbf{Incoming Connections:}
In the context of UDP peer-to-peer connectivity, an incoming connection occurs when one peer establishes communication with another peer by sending a UDP packet. The process of establishing an incoming connection typically involves the following steps:

\texttt{Peer Discovery:}
Peers within a UDP peer-to-peer network employ various mechanisms to discover and identify each other. This can include techniques such as broadcasting, multicasting, or using a centralized server for peer coordination.

\texttt{Packet Exchange:}
Once a peer has discovered another peer, it can initiate communication by sending a UDP packet to the target peer's IP address and port number. The packet may contain information about the requesting peer's identity, the desired data, or any other relevant details.

\texttt{Response Handling:}
The receiving peer processes the incoming packet and formulates an appropriate response. The response may contain the requested data, acknowledgement, or other necessary information. This two-way communication enables the establishment of an incoming connection between the two peers.

UDP peer-to-peer connectivity allows for decentralized and efficient communication between peers, making it suitable for various applications such as file sharing, voice and video streaming, and online gaming.
 
\subsection{Network Address Translators}
\label{nats}
\begin{figure}[h]
\includegraphics[width=8cm]{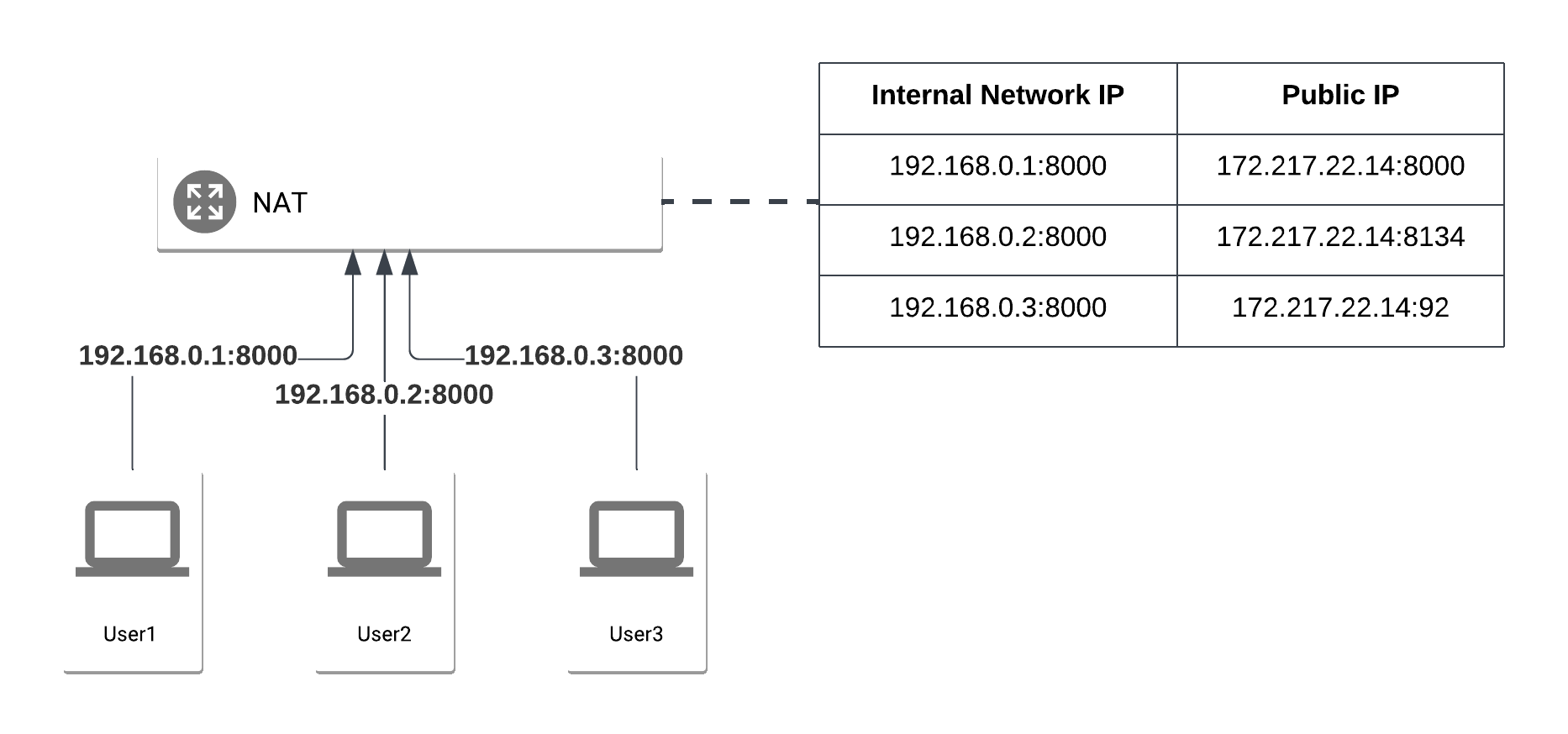}
\caption{A NAT setup}
\label{figure:nat}
\end{figure}

A network Address Translator (NAT) is a piece of hardware or software that holds a table of pairs of local and globally unique IP addresses. Locally IP addresses within the stub domain are not globally unique, so they cannot be used to route packets on the Internet. The packets will have the NAT's public IP address and when they are routed to it, the NAT will look up the local IP address that corresponds to the specific IP:Port pair the received packet holds. This method is used to address the problem of IP address depletion by basically ``bundling`` a LAN behind the NAT box and routing all packets using a single IP address \cite{the_ip_network_address_translator}. A potential home NAT setup is depicted in figure \ref{figure:nat}\par

However, it may not be a suitable long-term solution and could also lead to short-term problems. NATs face scalability issues since as tables scale there will be an overhead on the table lookup. Dynamically allocating IP:Port pairs also increase the probability of misaddressing since a delayed packet may end up on a collision since the NAT may have already re-allocated that pair. A crucial problem with NATs is that some applications cannot function properly or they break. Examples of these applications are video streaming, online games and P2P-dependent applications. \cite{the_ip_network_address_translator}.\par

According to the STUN protocol \cite{rfc3489}, there are four types of NATs, namely Full-cone NAT, Restricted-cone NAT, Port-restricted cone NAT and Symmetric NAT. These types fall into two categories according to RFC4787 \cite{rfc4787} namely the ``easy`` NATs which do Endpoint-Independent Mapping (EIM) and the ``hard`` NATs which do Endpoint-Dependent Mapping. EIM ensures the consistency of the external address and port pair if the request is coming from the same internal port.\par

According to Huawei \cite{qiaoqiao_2021} the specifications of these types of NATs are:\par

\textbf{Full-cone NAT} An \texttt{EIM} NAT where all requests coming from the same internal IP1:Port1 pair are mapped to the same public IP2:Port2 pair. On top of that, any host on the Internet can communicate with the host on the LAN by sending packets to the mapped public IP address and port. \par

\textbf{Restricted-cone NAT} An \texttt{EIM} NAT where similar to the Full-cone NAT an internal IP:Port pair will be mapped to the same external IP:Port pair. The difference with this NAT is that a host on the Internet can send packets to a machine behind the NAT only if that machine initiates the communication.\par

\textbf{Port-restricted cone NAT} An \texttt{EIM} NAT, similar to the Restricted-cone NAT but the restriction includes port numbers.

\textbf{Symmetric NAT} An \texttt{EDM} NAT where all requests coming from the same internal IP1:Port1 pair are mapped to the same public IP2:Port2 pair. The difference with this NAT is that it also takes into account the destination of the packet i.e. a request from internal IP1:Port1 to external IP2:Port2 will have a different mapping from a request IP1:Port1 to external IP3:Port3 thus two consecutive requests from the same internal pair but to different external hosts will lead to two different mappings.\par

\subsection{Carrier-Grade NATs}
While hiding local home networks behind NAT boxes did help with alleviating the problem of depletion of addresses for a while; this was not enough. Many ISPs worldwide are running out of IP addresses to allocate, thus they resulted in bundling different customers and areas together behind a NAT box i.e. a Carrier-Grade NAT (CGNAT). ISPs worldwide started rolling out CGNATs but by 2016 it had received very little empirical assessment\cite{cgnats_in_the_wild, hp_nat444}.
Some general problems of the kind that the everyday user might encounter, were identified in a test performed on a Turkish ISP \cite{tuskish_isp}. These are:
\begin{itemize}
    \item Users are unable to access remote desktops or cameras 
    \item Users cannot open a port on demand or cannot access it if it is already opened.
    \item Being in a weak CGNAT IP pool may affect the user's connection speed and it can also result in high ping
    \item Latency issues can occur due to the extra hop node
    \item Issues can occur with services allowing registration or log-in from only one IP in the case multiple users behind the same CGNAT attempt to access it (a testimony of this also exists on a Ziggo forum from a user who complained about not being able to access google.com or any META-owned site \cite{ziggo_user})
\end{itemize}\par

An assessment contributed to the RFC series of the Internet Engineering Task Force (IETF) identified various services where a CGNAT may cause them to break or degrade their performance \cite{rfc7021}. In general, the testing revealed that applications such as video streaming, video gaming and P2P file sharing are impacted by CGNAT.

The services that broke are \cite{rfc7021}: 
\begin{itemize}
    \item Several P2P applications like XBOX P2P gaming and SIP calls, which use using PJSIP client, failed in both the NAT444 and Dual-Stack Lite environments (PJSIP worked when clients used a registration server to initiate calls, given that the client inside the CGNAT initiated the traffic. FTP sessions to servers located behind two layers of NAT failed. When the CGNAT was bypassed and traffic only needed to flow through one layer of NAT, clients were able to connect). 
    \item Applications that did not first send outgoing traffic thus not opening an incoming port through the CGNAT hence being unable to launch 
    \item Applications that tried to open a particular fixed port through the CGNAT, which works for a single subscriber but not when multiple subscribers try to use the same application. 
    \item Multicast traffic was not able to flow through the CGNAT.
\end{itemize}

The services whose performance was impacted are \cite{rfc7021}: 
\begin{itemize}
    \item Large file transfers initiated on the same (home) network
    \item Multiple video streaming sessions initiated in the same (home) network
    \item Sometimes video streaming like Silverlight and Netflix would exhibit a slowdown in single sessions as soon as a second session was established (router dependent issue) ---routers that support DSLite did not slow down on single-session video streaming when a second is established 
\end{itemize}

RFC7021 \cite{rfc7021} identified some additional problems, which are: 
\begin{itemize}
    \item Loss of geolocation information, something that mainly affects applications that require the precise location of the user and not an approximation of it, since the exact location becomes impossible to get and the only available one might be the location of the CGNAT box.
    \item Lawful Intercept/Abuse Response, explained in section \ref{legal-implications}
    \item Harder for security testers to launch anti-spoofing attacks.
\end{itemize}

\begin{figure*}[h]
\includegraphics[width=\textwidth]{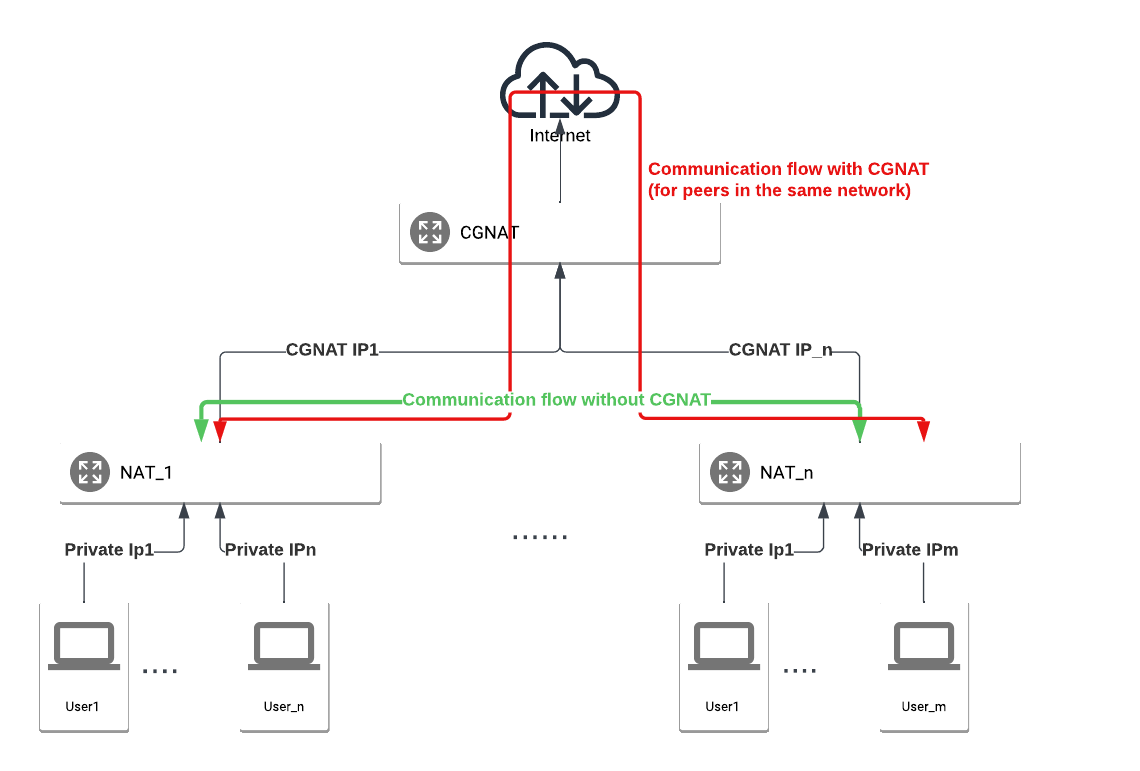}
\caption{Communication flow behind a CGNAT}
\label{figure:cgnat}
\end{figure*}

Much research was performed on the problems that NATs and CGNATs cause. For example, the thesis of Fredrik Thernelius \cite{thernelius_2000} identified problems occurring with the Session Initiation Protocol (SIP) when the session is not initiated from the inside of the NAT since otherwise extra steps will be required to find the address to the right internal host. Victor Paulsamy et al. \cite{paulsamy_chatterjee_2003} identified problems with NAT Firewalls when the Voice over Internet Protocol (VoIP) is used. \par

CGNATs often fail on peer-to-peer communication from users behind the same CGNAT as explained in section \ref{penetrating-nat}. The flow of communication is shown in figure \ref{figure:cgnat} with green being the intended communication flow and red being the actual communication flow.

\section{Impact of NATs on P2P protocols}
\label{impact-on-p2p}
The main interest of this study is how NATs, specifically CGNATs are affecting P2P communications since the majority of customers are protected by several levels of NAT. At the same time, data centre nodes may be hidden behind NAT for security or virtualization. The utilization of containerized deployments is exacerbating the situation since every communication between peers necessitates a mechanism to navigate NATs, or else operations will be impacted \cite{ietf_draft_trustchain}. This problem comes in two-fold since NATs sometimes act as Firewalls meant to block incoming traffic from entering the LAN unless those packets are a response to a communication established from the inside.\par

J.J.D Mol et al.\cite{mol_pouwelse_epema_sips_2008} established in their 2008 research on fairness for BitTorrent users chain, that peers that are behind firewalls have more difficulty obtaining a fair sharing ratio; thus they concluded the need for puncturing NAT or using a static IP address to optimize the performance of the network.\par

This problem of clients behind NATs being unfavored by the BitTorrent protocol can be highly observed in networks like Tribler --- a BitTorrent \cite{bittorent} based open source project that extends the protocol by adding features such as video on demand and live streaming while remaining fully backwards compatible \cite{tribler}--- because the vast majority of the peers are mobile devices. The vast majority of mobile phones are behind NATs since there is no option for a static IP address on a mobile device, and cellular networks use carrier-grade NATs.\par

As noted by J. Pouwelse et al. \cite{measurement_of_nat_and_firewalls}, most consumers are behind several layers of NATs thus the new implementations/versions of P2P-based networks should take this into account. According to the same authors, to build effective P2P networks, the network designers need to consider that many users are behind non-specially configured NATs and firewalls in their home setups (potentially also CGNATs). To date, no elegant solution has been found for P2P TCP-based communication through NAT/firewall, which means that the network needs to be UDP-based to allow for NAT traversals.\par

\section{Penetrating a NAT}
\label{penetrating-nat}
As established in Section \ref{impact-on-p2p}, most users will have some troubles with P2P communication when they reside behind a NAT and/or a firewall since P2P networks assume that all nodes in the network are connectable. For UDP-based protocols, the UDP hole-punching technique has been proposed to overcome this problem. The idea is since the NAT will block packets that are incoming from connections that have not been mapped yet, then the internal endpoint must send some packet first to the external remote endpoint thus creating a ``hole`` in the NAT or firewall through which then communication can then proceed \cite{halkes_pouwelse_2011, ford2005peer, ietf_draft_trustchain}. Note that TCP puncturing is also possible but adds extra levels of complexity thus if TCP is necessary due to a stream-oriented connection, the user should consider switching to QUIC and then adapt the UDP-puncturing techniques to work with QUIC \cite{quic}. Different steps (which will be explained in this chapter) are required based on the type and amount of NATs/Firewalls and whether one or both users are behind a NAT/Firewall.\par

As mentioned in Section \ref{nats}, NATs are divided into ``easy`` and ``hard`` with easy being the ones which are relatively easier to penetrate. The algorithm to penetrate NATs using the Birthday Paradox \cite{swadling2019birthday} is derived from a blog post series by David Anderson \cite{anderson_2020, anderson_2022}. Note that throughout this text the terms NAT traversal, NAT penetration, puncturing, and UDP hole-punching will be used interchangeably.\par

Before going to the algorithm, it's good to note that most NATs include a stateful firewall which is a piece of software/hardware that recalls the packets previously encountered and applies that information when determining how to handle new packets that arrive.\par

Starting, to perform NAT traversal there are two requirements, that is \textit{the communication protocol between the two peers need to be UDP based} and \textit{the peer(s) behind the NAT needs to have direct control over the network socket that is sending and receiving network packets}. To bypass the stateful firewall using UDP is straightforward: the firewall permits an incoming UDP packet only if it has previously detected a corresponding outbound packet. Hence the computer located behind the firewall must be the initiator of the connections. However, a problem arises when two peers located behind firewalls wish to communicate directly (since the firewalls are now "facing each other"). No one can make the first move because each side is waiting for the other to take the initiative.\par

\begin{figure*}[h]
    \begin{center}
       \includegraphics[width=15cm]{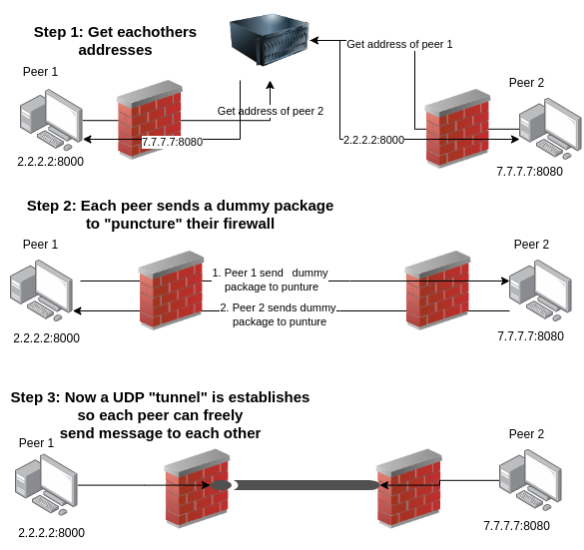} 
    \end{center} 
    \caption{Two machines behind Firewalls using a synchronizer server to find each other's IP address and then puncture their firewalls to establish a communication channel between each other}
    \label{fig:firewalls_with_synch_server}
\end{figure*}

The solution to the double firewall problem comes from the observation that the firewall rule says that packets must flow out before packets can flow in, \texttt{but they don't necessarily have to be related to each other}. As long as the incoming packet has the expected source and destination then any packet can be a response to the outgoing one. Thus to traverse multiple stateful firewalls, there needs to be some information shared in advance, i.e. the IP:port that each peer is using, which can be either manually configured --- something that does not scale quickly--- or the peers can use a coordination server to keep the IP:port information synchronized securely and flexibly as in figure \ref{fig:firewalls_with_synch_server}. Note that this technique requires precise timing, roughly the second step (of initiating the puncture), needs to be done at the same time thus a clock should be used.\par

There are three possible scenarios, both peers being behind EIM-based NATs, one being behind an EIM-based NAT and the other behind an EDM-based NAT or both peers being behind EDM-based NATs.

\subsection{Peers that are both behind an EIM-based NAT}
Now when both peers are behind an EIM-based NAT things get harder, since the peers don't know their external IP addresses ---strictly speaking, there is no external IP:port until the other peer sends packets since NAT mappings are only created when outbound packets are required to flow towards the Internet. Both need to initiate the communication first, but no one knows who to send the packet to. This problem of one not knowing their IP address is what the Session Traversal Utilities for NAT (STUN) is aiming to solve. The idea behind it is that when a client behind a NAT communicates with a server on the Internet, the server sees the public IP address of the client, not the intranet one. Thus the server can reply to the client with the IP:port that the server saw when it received the message \cite{rfc3489}. Note that depending on the NAT type, the IP:port that STUN ``sees`` is not always the same as the one the whole Internet sees. EDM-based NATs as explained in section \ref{nats} create a different mapping for every single destination.\par

\subsection{Peers where one is behind an EIM-based NAT and the other behind an EDM-based NAT}
If one peer is behind an EDM-based NAT that means that the NAT is opening a different port for each communication. Since there are 65535 different ports \cite{cloudflare_learning}, if the peer behind the EIM-based port attempted to brute force it, assuming median Internet speed in the Netherlands which at the time of writing is $\approx$ 128 Mbps download and $\approx$ 40 Mbps upload \cite{speedtest} the maximum packets per second that a machine can send is $\approx$ 57000 packets per second (assuming minimum UDP packet size of 8 bytes and minimum ethernet packet overhead of 84 bytes \cite{ethernet}), meaning that it can brute force the NAT mapping in little over 1s ($65535 \div 57000$).\par

Problems arise when the Internet upload speed is low, the NAT/Firewall has some rules to block brute-forcing or both of the peers are behind an EDM-based NAT.\par

\subsection{Peers where both are behind an EDM-based NAT}
Assuming that someone would try to brute force between two peers that are both behind EDM-based NATs. Since the initiator of the brute force attack does not know their mapping either, this means that now there are $65535^2 = 4294836225$ possible combinations on a \{source port, destination port\} pair. 
Assuming the average Internet connection in the Netherlands, a brute-force attack would take roughly 21 hours ($4294836225 \div 57000$).\par

To improve on this, one can use the Birthday Paradox to achieve a collision in significantly less time.\par

The \textbf{Birthday Paradox} is a problem in probability theory that involves determining the likelihood of at least two individuals sharing the same birthday among a group of n randomly selected people. Surprisingly, the paradox reveals that only 23 people are needed to reach a 50\% probability of shared birthdays. This may seem counter-intuitive, but the reasoning behind this is that every possible pair of individuals within the group will be compared. Therefore, with 253 pairs to consider (calculated by $\frac{23 \cdot 22}{2}$), which is more than half the number of days in a year, it becomes easier to understand why this result holds \cite{swadling2019birthday}. This paradox also has practical applications, such as the "birthday attack," which uses this probabilistic model to reduce the complexity of finding a collision. In this case trying to find a collision of two pairs of IP:port.\par

From the calculations performed in \cite{anderson_2020} one can get a 50\% success rate of double IP:port collision after sending $\approx$ 54000 packets, something that can be achieved in under a second, assuming an average Netherlands connection. To achieve a 99.9\% success rate $\approx$ 170000 packets are needed \cite{swadling2019birthday}, where assuming the same connection it can be achieved in little over 3 seconds. This is a tremendous improvement from the 21 hours needed without using a birthday attack. Note that a birthday attack could also be used in the scenario of one peer behind an EIM-based NAT and the other behind an EDM-based NAT to achieve an almost instant puncture with a 99.9\% probability.\par

Although this improvement is great in theory, the problem comes when considering that some routers max out at 64000 active connections (for example Juniper SRX 300). A saviour bet is that 54000 probes from each side lead to a 50\% probability of puncture. In the case that a puncture has yet to be achieved, more probes can be performed with the hope that the routers will behave gracefully in the case of overloading.

The code associated with the algorithm above can be found on this GitHub page \cite{danderson}.\par

\begin{figure*}
    \includegraphics[width=\textwidth]{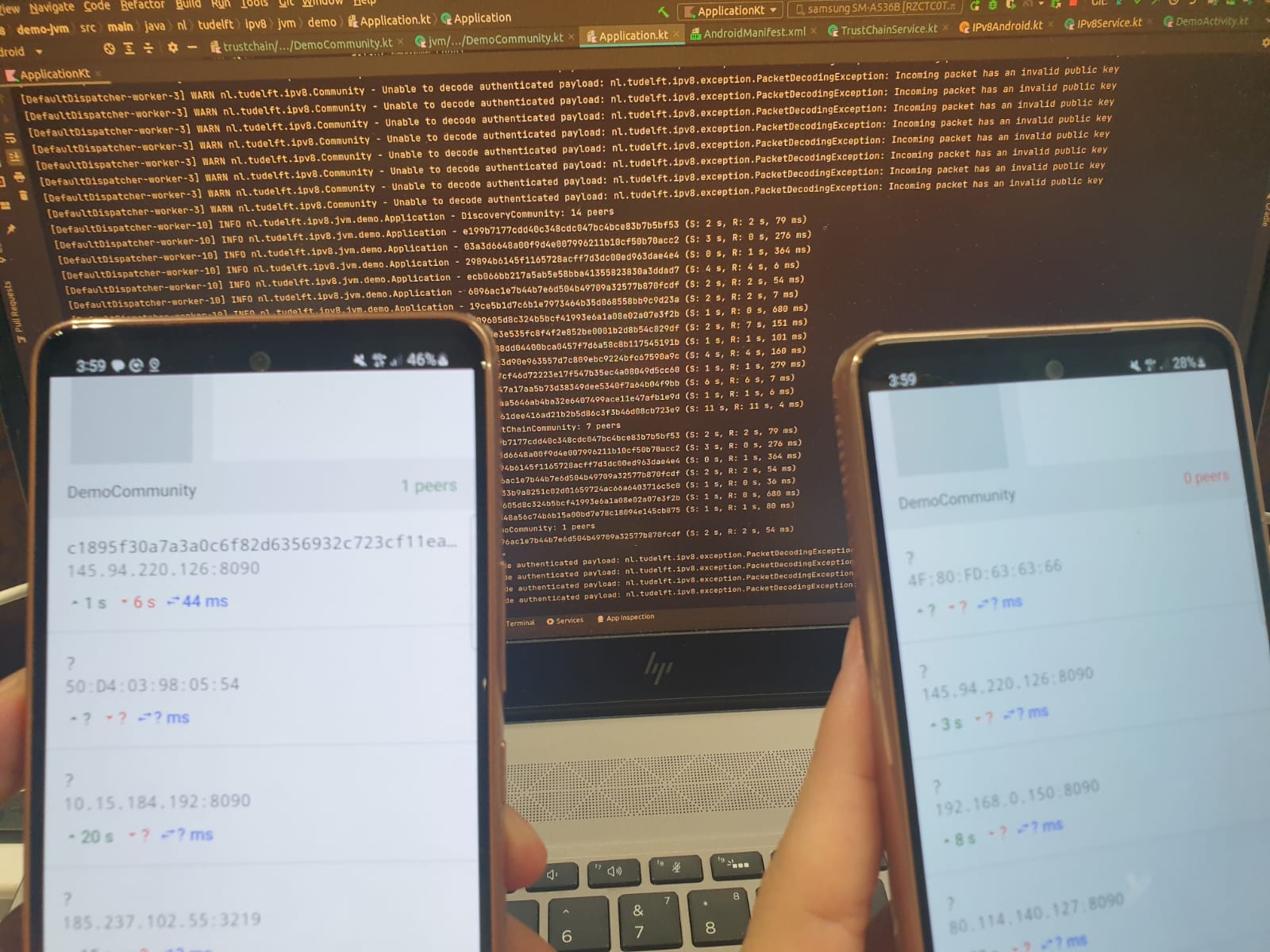}{\caption{The setup used}\label{fig:experimental-setup}}
\end{figure*}

\begin{figure}[H]
    \includegraphics[width=\textwidth]{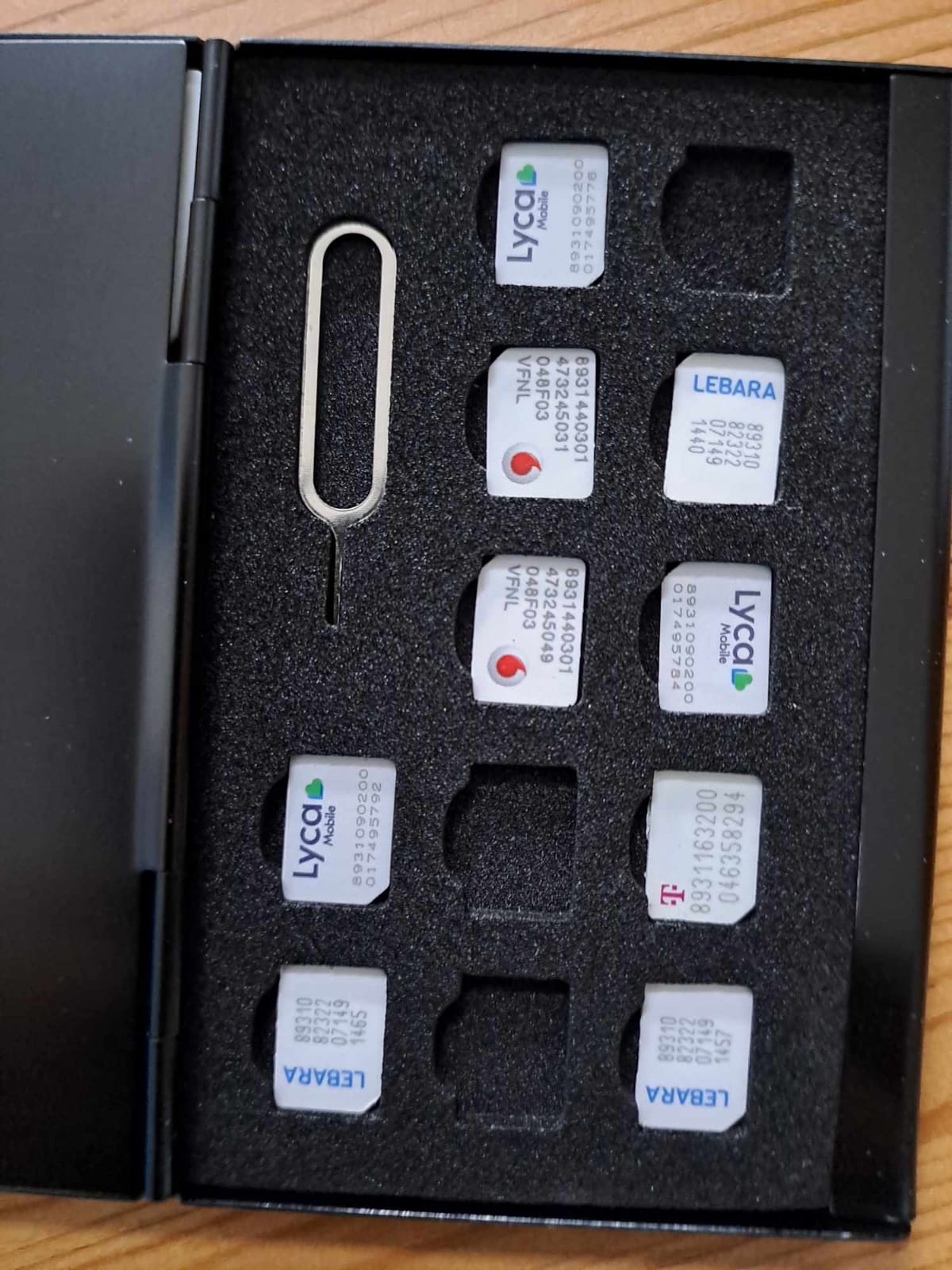}{\caption{The sims used for testing}\label{fig:experimental-setup2}}

\end{figure}

\subsection{Port Mapping Protocols}
\label{port-mapping-protocols}
There exist three protocols for partially manipulating port maps. Something like asking the NATs to allow more stuff in, so we don't have to result in brute force or birthday attacks. There are three protocols to do this and they will be explained in this subsection.

\textbf{Universal Plug`n`Play Internet Gateway Device (UPnP IGD)} is the oldest of the three, developed in the late 1990s and uses old technologies like XML, SOAP and multi-cast HTTP over UDP. It allows the user to perform a request to map ports in their UPnP IGD-enabled NAT. One should keep in mind that is hard to implement correctly and securely \cite{rfc6970}.\par

\textbf{NAT Port Mapping Protocol (NAT-PMP)}, a protocol developed by Apple as a competitor of UPnP IGD was designed only to perform port forwarding while being much easier to implement both on clients and on NAT devices \cite{rfc6886}.\par

\textbf{Port Control Protocol (PCP)} also known as NAT-PMPv2 is similar to NAT-PMP but also support IPv6 \cite{rfc6887}.

Having these protocols in mind one can try to look up any of these three protocols on their local default gateway and request a public port mapping thus further simplifying the connectivity between hosts behind NATs; although it is common that these protocols are disabled.\par

To decide on what technique to use at any time, be it brute force, birthday paradox, port mapping etc. there is the Interactive Connectivity Establishment (ICE) developed in 2010 \cite{rfc8445}.

\textbf{Interactive Connectivity Establishment (ICE)} simply put is an algorithm where one tries every technique at once and picks the best technique that works. According to IETF,  the ICE protocol is a NAT traversal technique, a multi-homed address selection technique and a dual-stack address selection technique that works by including multiple IP addresses and ports in both the request and response messages of a connectivity establishment transaction without making any assumption about the network topology \cite{rfc8445}.\par

\subsection{CGNATs}
Before the rollout of CGNATs, users could bypass their NATs by using any port mapping protocol to configure port forwarding on their home routers. Unfortunately, the ISPs' CGNATs are not reconfigurable.\par
Fortunately, though the existence of a CGNAT throughout the routing path will not require any significant changes since it is practically a double NAT so the same algorithm can be used since the only NAT that should concern the user is the last one from the Internet. Still, one should expect that more time will be required to crack them since there will be significantly more combinations to test.\par

A problem arises when both peers are behind the same CGNAT. This is a problem since STUN won't work since the STUN server will be outside of the intranet and will see the ``middle`` network --- since CGNAT effectively develops a ``mini`` Internet for the  devices connected to it---every time a ``what is my address`` request is performed.\par

In case hairpinning is supported by the CGNAT we can try to use that. Hairpinning is when both internal peers use a STUN server to get their external IP address and then the other peer just sends the packets to that IP address hoping that they will go through. Though it is not always the case that hairpinning is supported by the CGNATs \cite{rfc4787}.\par

In the case that hairpinning fails, then the user's only option left is relaying.\par

\subsection{pwnat}
A promising solution named \texttt{pwnat} rolled out in 2010, claimed autonomous NAT traversal without the need of any third parties by using fake ICMP messages to initially contact the NATed peer.\par

While pwnat is still in operation, it can mainly work in cases where only one of the parties is behind a NAT, not both, according to the homepage of pwnat claiming that the client has somehow leaned the current external IP address of the peer's NAT \cite{muller2010autonomous}. Another problem comes when the server transmits ICMP\footnote{Internet Control Message Protocol} echo request packets to the specified address, in accordance with RFC 3022 \cite{rfc3022}, the identifier field within the ICMP echo request header is uniquely associated with a query identifier linked to the registered IP address by NATs. This association enables NAT-S to subsequently direct ICMP Echo Replies with the same query ID back to the sender. Consequently, it becomes necessary to modify the ICMP header within ICMP Query packets to replace the query ID and ICMP header checksum, as outlined in the RFC 3022 ICMP error packet modification section.\par

In the context of a NAPT\footnote{Network Address and Port Translation} setup, if the IP message contained within ICMP happens to be a TCP, UDP, or ICMP Query packet, you must also adjust the appropriate TU port number within the TCP/UDP header or the Query Identifier field within the ICMP Query header. However, the client is unaware of the external query ID (the code in pwnat uses 0 as the identifier for the original request). As a result, it sends an ICMP Time Exceeded packet to the server. Even if this packet manages to reach NATs in front of the server, NATs are unable to locate the active mapping for the embedded packet, leading to the likelihood that most NAT implementations will discard it.

\section{Reproducing results from literature}

A simple app running IPv8\footnote{https://github.com/Tribler/kotlin-ipv8} was developed to test the penetration rate of NATs on mobile on various Dutch telecom providers. Four different SIM cards were used i.e. Lyca\footnote{https://lycamobile.nl/en/}, Lebara\footnote{https://mobile.lebara.com/nl/en}, T-Mobile\footnote{https://www.t-mobile.nl/} and Vodafone\footnote{https://www.vodafone.nl/}.\par

\begin{figure*}[h!]
\includegraphics[width=\textwidth]{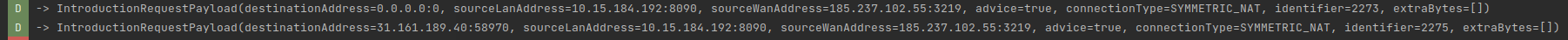}
\caption{IPv8 reporting failure on penetrating Lyca due to Symmetric NAT}
\label{figure:lyca}
\end{figure*}

The goal of the app was to determine whether IPv8 could make two phones each on a different carrier's 5G running Kotlin-ipv8 and a computer running the JVM version of IPv8 discover each other and communicate by penetrating potential NATs that are in the way as can be seen in figures \ref{fig:experimental-setup} and \ref{fig:experimental-setup2} . The experiment failed on Lyca since IPv8 was unable to penetrate the Symmetric NAT that Lyca has in place as can be seen in figure \ref{figure:lyca}.\par

\begin{figure*}
  \begin{floatrow}
    \ffigbox[\FBwidth]{\includegraphics[width=\textwidth]{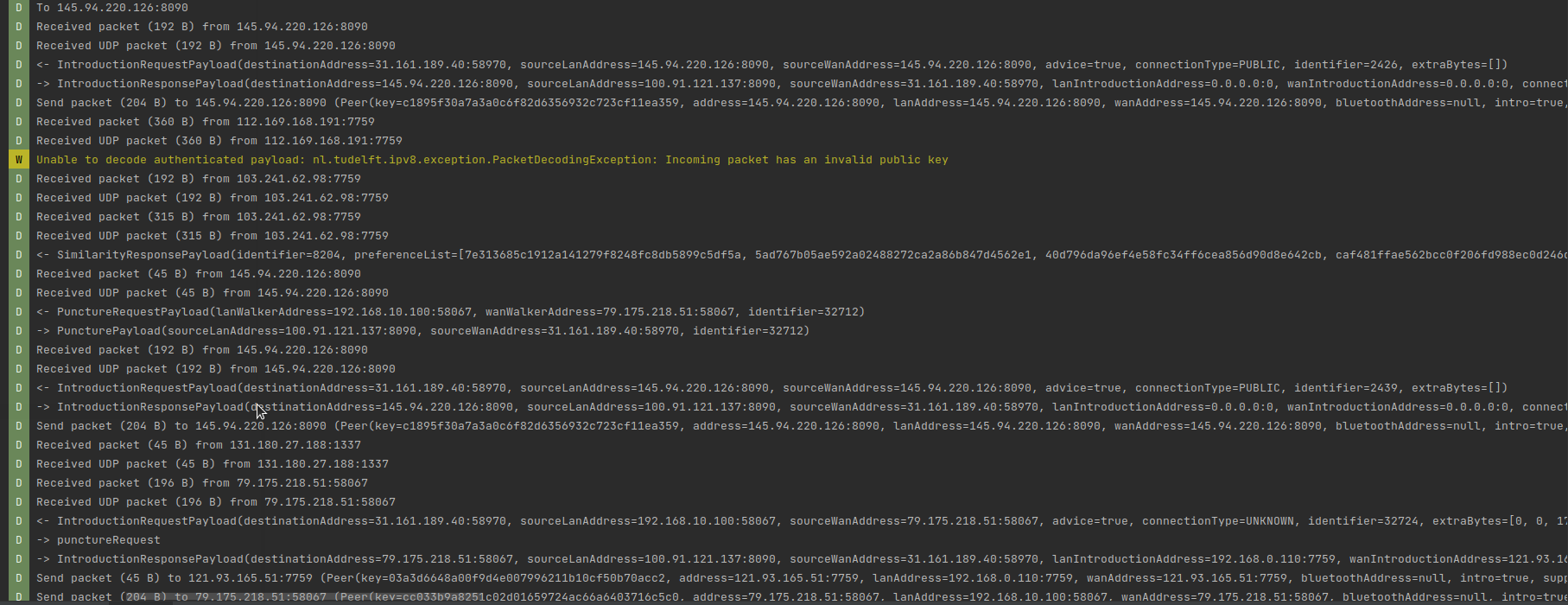}}{\caption{Lebara successfully communicating with Lebara and the JVM}\label{figure:successful-comm}}
  \end{floatrow}
  \begin{floatrow}
    \ffigbox[\FBwidth]{\includegraphics[width=\textwidth]{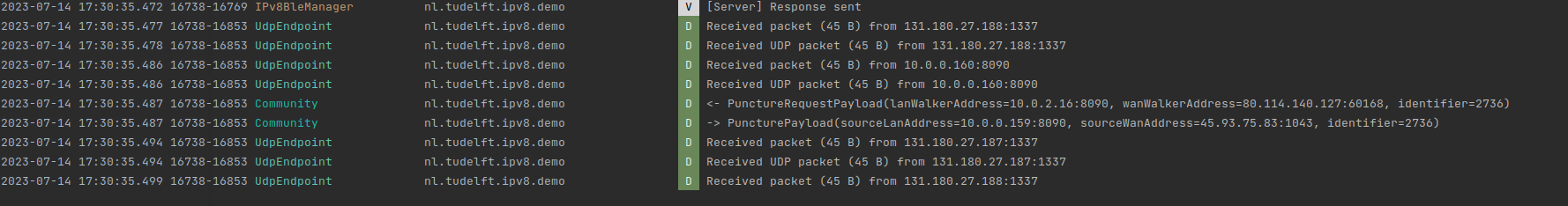}}{\caption{Logs of the successful puncturing}}
  \end{floatrow}
  \begin{floatrow}
    \ffigbox[\FBwidth]{\includegraphics[width=\textwidth]{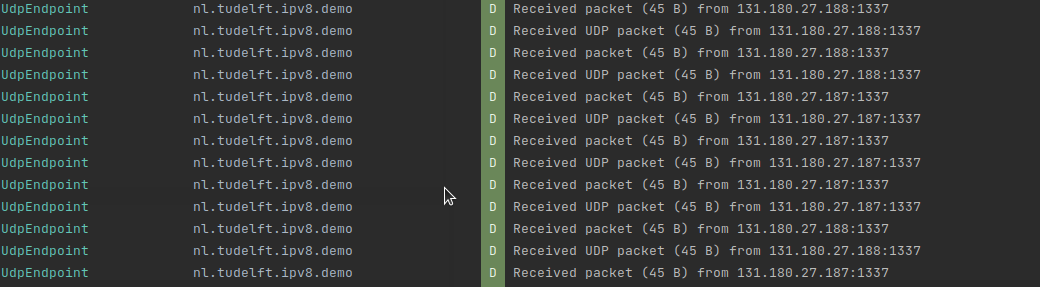}}{\caption{Packets send and successfully received}}
  \end{floatrow}
\end{figure*}

The other 3 carriers achieved communication both with each other and with the computer (as can be seen in table \ref{tab:communications} and figure \ref{figure:successful-comm}), with IPv8 reporting that it managed to exchange the preset UDP messages between the devices while also penetrating the NATs that were in place and keeping the hole alive throughout the whole duration of the experiment. This represents a 75\% success rate of NAT penetration across the four different Dutch carriers tested.\par 

\begin{table}[]
\begin{tabular}{lllll}
\cellcolor[HTML]{C0C0C0} & T-Mobile                           & Lebara                             & Lyca Mobile                    & Vodafone                           \\
T-Mobile                 & \cellcolor[HTML]{34FF34}\checkmark & \cellcolor[HTML]{34FF34}\checkmark & \cellcolor[HTML]{FE0000}$\times$ & \cellcolor[HTML]{34FF34}\checkmark \\
Lebara                   & \cellcolor[HTML]{34FF34}\checkmark & \cellcolor[HTML]{34FF34}\checkmark & \cellcolor[HTML]{FE0000}$\times$ & \cellcolor[HTML]{34FF34}\checkmark \\
Lyca                     & \cellcolor[HTML]{FE0000}$\times$     & \cellcolor[HTML]{FE0000}$\times$     & \cellcolor[HTML]{FE0000}$\times$ & \cellcolor[HTML]{FE0000}$\times$     \\
Vodafone                 & \cellcolor[HTML]{34FF34}\checkmark & \cellcolor[HTML]{34FF34}\checkmark & \cellcolor[HTML]{FE0000}$\times$ & \cellcolor[HTML]{34FF34}\checkmark
\end{tabular}
\caption{Carriers that succeeded in transmitting a package to another carrier}
\label{tab:communications}
\end{table}

There was no packet loss observed between the three carriers, but there were some instances where the received packets had invalid public keys, thus they could not be decoded.\par

\section{Legal Implications of Carrier-Grade NATs}
\label{legal-implications}
As already discussed, CGNATs may crush or slow down some applications, but they may impact law enforcement investigations into online crime.\par

Part of these investigations is to gather intelligence which requires knowledge of Internet communications. This was much easier before NATs since ISPs would just store a table of IP addresses, who used them and the period of that. ISPs are generally required to store this information under the ``data retention`` law.\par

With the rollout of CGNATs, this format of data retention is not sufficient anymore since one does not have their IP address allocated. The new format should be, the IP address allocated, the IP address used and the port number along with a timestamp, but this is very inefficient due to the huge amount of data that this format will generate. Assuming a large ISP of 25 million customers, and users averaging 33 thousand connections a day, this averages to a log file of about 425TBytes. This is highly inefficient and significantly slow to query\cite{huston_2013}.\par

Politicians, law enforcement agencies and ISPs are collaborating to format the new laws around data retention to allow law enforcement to find the perpetrator of a crime while not overloading the ISPs' data centres. This should be done with great care since when multiple people are sharing a single IP address when a crime is committed there should be a way to identify the exact source of the crime to avoid wrongly investigating innocent individuals who happen to share an IP address with a criminal \cite{europol_2017}.

\section{A better Approach than Carrier-Grade-NAT}
The most obvious solution to removing CGNATs is a full rollout of IPv6, completely replacing IPv4. Although this is the theoretical eventual goal, it won't be realized soon. As of April $23^{rd}$ 2023 --- 11 years after the launch of IPv6---, Google statistics show 41.93\% of their users having adopted IPv6 \cite{google_2023} thus it is fair to assume that there is a long way until a complete IPv6 adoption by the Internet.\par

Since there is no specific day that IPv4 support will come to an end and a full transition to IPv6 will be rolled out, some nodes on the internet are currently only running on IPv4 and some on IPv6. A core value of the internet is that any node can communicate with any other node, thus a seamless translation mechanism is required between these nodes \cite{durand-ngtrans-nat64-nat46-00}.\par
Enter NAT46 and NAT64. These are network address translators but their purpose is to translate from IPv4 to IPv6 and vice-versa. The need for these NATs is currently unavoidable at this state but it also introduces the same problems as regular NATs. \par

Until IPv6 is completely rolled out, different researchers have proposed solutions to depleting IPv4 addresses that do not involve CGNATs. One of these suggestions came from O. Maennel et al. \cite{maennel2008better} where they suggested extending the IPv4  address space by "stealling" bits from the port number in TCP/UDP ---thus still allowing users to utilize a single IPv4 address.\par

Their suggestion called extended addressing, is to limit the port addressing to 6 bits, an action that will increase the address space by 10 bits, thus multiplexing 1024 additional users per existing IP address.\par

This suggestion will reduce the number of fixed ranges of ports that applications can use. Still, the need for IP addresses is stronger than the need to run thousands of applications on a single host, something that broadband consumers are not anticipated to do.\par

\section{Discussion}
This paper highlights the issues associated with the wide adoption of NATs particularly CGNATs and how they affect different applications while also raising legal and societal issues. The ideal solution is full adoption of IPv6, which is not coming anytime soon thus NATs are unlikely to go away any time soon.\par 

Multiple different solutions were presented in the paper, including NAT puncturing to allow the smooth working of peer-to-peer protocols to an adaptation of IPv4 where fewer bits from the port number will be used to extend the address part of IPv4, thus increasing the available address space. Each solution is good for different scenarios and it also depends on whether the user can implement these techniques.\par

While most of the solutions mentioned throughout this survey are mainly targeted at servers and PCs, there is research performed by Diego Madariaga et al. \cite{madariaga_torrealba_madariaga_bustos} that measures the quality of service provided by mobile operators using ICMP\footnote{\url{https://en.wikipedia.org/wiki/Internet_Control_Message_Protocol}}
messages to measure the Round-Trip time, jitter and packet loss of a mobile's operator network. This research is of high interest since the eventual goal of this survey is to develop state-of-the-art NAT penetration on mobile phone networks, thus the insights from this paper i.e. the packet loss rate can be used to derive information on the NAT (types) used by the mobile networks.

Fortunately, according to Bryan Ford et al. \cite{ford2005peer}, as NAT vendors become more aware of the requirements of significant P2P applications like Voice-over-Internet protocol and online gaming protocols, they will probably enhance their support for hole punching in the coming times.\par



\end{document}